# Atomic origin for hydrogenation promoted bulk oxygen vacancies removal in vanadium dioxide

Bowen Li[a,+], Min Hu[b,+], Hui Ren[a], Changlong Hu[a], Liang Li[a], Guozhen Zhang[b], Jun Jiang[b,*], Chongwen Zou[a,*]

[a] National Synchrotron Radiation Laboratory, University of Science and Technology of China, Hefei, 230029, China

[b] Hefei National Laboratory for Physical Sciences at the Microscale, Collaborative Innovation Center of Chemistry for Energy Materials, CAS Center for Excellence in Nanoscience, School of Chemistry and Materials Science, University of Science and Technology of China, Hefei, Anhui 230026, P. R. China

+These two authors contributed equally to this paper.

*Corresponding Authors: jiangj1@ustc.edu.cn ; czou@ustc.edu.cn

**Abstract**

Oxygen vacancies ($V_O$), a common type of point defects in metal oxides materials, play important roles on the physical and chemical properties. To obtain stoichiometric oxide crystal, the pre-existing $V_O$ is always removed via careful post-annealing treatment at high temperature in air or oxygen atmosphere. However, the annealing conditions is difficult to control and the removal of $V_O$ in bulk phase is restrained due to high energy barrier of $V_O$ migration. Here, we selected $VO_2$ crystal film as the model system and developed an alternative annealing treatment aided by controllable hydrogen doping, which can realizes effective removal of $V_O$ defects in $VO_{2-\delta}$ crystal at lower temperature. This finding is attributed to the hydrogenation accelerated oxygen vacancies recovery in $VO_{2-\delta}$ crystal. Theoretical calculations revealed that the H-doping induced electrons are prone to accumulate around the oxygen defects in $VO_{2-\delta}$ film, which facilitates the diffusion of $V_O$ and thus makes it easier to be removed. The methodology is expected to be applied to other metal oxides for oxygen-related point defects control.

# Introduction

As a strongly correlated oxide material, vanadium dioxide ($VO_2$) undergoes a typical metal-insulator transition (MIT) with the critical temperature around 340 K [1,2]. During the phase transition, the conductivity shows a sharp change up to 4-orders-of-magnitude and the infrared transmittance demonstrates a pronounced switching effect, which changes from high reflection state to high transmission state. These unique phase transition makes $VO_2$ a promising candidate for various applications, such as "smart windows", optical switches, phase transition transistors and memristor neurons [3-6].

However, the variability of valence states of vanadium and multiple phase structures of $VO_2$ will easily induce point defects. Especially, the oxygen vacancies ($V_O$) in crystal lattice may degraded the phase transition behavior and undercut its potential of applications [7-9]. Removal of these oxygen vacancies typically requires high-temperature annealing in air or oxygen atmosphere. Unfortunately, it is quite difficult to precisely control the temperature, pressure and time for annealing process, because of multi-state nature of V ions [10-13]. Further, $V_O$ in deeper layer of $VO_2$ crystal is reluctant to diffuse, making it difficult to be completely removed. Thus, it is highly desirable to decrease the diffusion barrier of $V_O$ and achieve a controllable removal, which would be useful for stoichiometric $VO_2$ material preparation.

Kamiya et al. [14] and Kim et al. [15] respectively reported that the $H_2$ plasma treatment could decrease the oxygen vacancies and increase the surface oxygen species on amorphous indium–gallium–zinc–oxide (a-IGZO) thin-film transistor. Similarly, Chen et al. [16] demonstrated the hydrogen would remarkably enhance the performance of InGaZnO field-effect transistors and observed the reduction of oxygen vacancies. Luo et al. [13] observed an increasing of oxidation rate of Ni-Cr alloy in water vapor using in-situ TEM and proposed a hypothesis that the protons will occupy the interstitial positions in oxide crystal lattice and lower the vacancy formation energy. All of these work suggest the interaction between hydrogen atoms and point defects in oxide crystal lattice may play a role in the oxygen removal. While, the precise mechanism and the dynamics of vacancies diffusion in oxides by external hydrogen doping were still unclear.

In this work, we selected VO$_2$ crystal film as the model system of interest, because its MIT behavior was very sensitive to the oxygen vacancies. Epitaxial VO$_2$ film was firstly synthesized by MBE method, then annealed in vacuum chamber to form the preferred VO$_{2-\delta}$ thin film with high concentration of V$_O$. We found that the metallic VO$_{2-\delta}$ thin film was quite stable even heated in the air at 200 °C. After a treatment of hydrogen doping, the same annealing process is able to completely remove the oxygen-vacancies and produce a pure VO$_2$ film. This observation was attributed to the hydrogenation accelerated oxygen vacancies recovery in VO$_{2-\delta}$ crystal. The following theoretical calculations revealed that the electrons are prone to accumulate around the oxygen defects in VO$_{2-\delta}$ film after hydrogen doping, which facilitates the diffusion of V$_O$ and thus makes it easier to be removed.

**Results**

Fig. 1(a) shows transformation between the M-VO$_2$ insulator and the oxygen vacancies involved VO$_{2-\delta}$. The pristine M-VO$_2$ crystal has pronounced MIT behavior, while if some oxygen vacancies was produced (See Fig. 1S), the VO$_{2-\delta}$ will be metallic state at room temperature. For H-doped VO$_2$ crystal, the insulator-metallic-insulator phase transitions will be observed upon the H-doping concentration (Fig. 1b), which has been reported by the recent literatures [6,17]. For VO$_{2-\delta}$ metallic film, it is quite stable in air. It was difficult to transfer to stoichiometric VO$_2$ even heated up to 200°C. While if doping some H atoms into VO$_{2-\delta}$ film, the obtained H$_x$VO$_{2-\delta}$ would change to pure VO$_2$ effectively under the same heat conditions (Fig. 1c).

The scheme shown in Fig.1c is clearly confirmed by the experimental observation in Fig. 1d, which shows the electrical resistance as the function of annealing time for VO$_{2-\delta}$, H$_x$VO$_{2-\delta}$ and HVO$_2$ samples by heating in air at 200°C. It can be observed that the metallic VO$_{2-\delta}$ film is quite stable after heating in air for more than 350min. The fully hydrogenated HVO$_2$ film transfers quickly from insulator state (~1.5MΩ) to metallic state (~1.0KΩ), and then goes to insulator state (~1.0MΩ) within several minutes. These quick transformation for fully hydrogenated HVO$_2$ film should be mainly due to the fast escaping of H atoms in crystal lattice. While for H-doped $VO_{2-\delta}$ film (or $H_xVO_{2-\delta}$), the phase transformation upon the heating treatment is quite different from the $VO_{2-\delta}$ sample. It is clear that the H-doped $VO_{2-\delta}$ film can be recovered to insulator $VO_2$ after the heat treatment. The R-T curves shown in Fig. 4e further confirm that the $VO_{2-\delta}$ film still keeps its metallic property, while the $H_xVO_{2-\delta}$ film will recover to M-$VO_2$ with hysteresis loop after the same heat treatment.

Fig.2 (a) and (b) showed the XRD and Raman spectra for the $VO_{2-\delta}$ film and the 5-hour annealed $VO_{2-\delta}$ and $H_xVO_{2-\delta}$ film samples. For comparison, the measurements for pristine $VO_2$ film were also plotted (See Fig. 2S). The XRD peak at $2\theta=39.8^o$ was from the (020) diffraction of pristine $VO_2$ film. For $VO_{2-\delta}$ film, the XRD peak shift to $2\theta=39.5^o$. If annealing this $VO_{2-\delta}$ film in air at 200$^o$C for five hours, the XRD peak shifted a little towards higher angle direction. While if doping some hydrogen atoms into $VO_{2-\delta}$ film ($H_xVO_{2-\delta}$), and then the $H_xVO_{2-\delta}$ sample was treated under the same condition, the XRD peak shifted to the same $2\theta$ angle as the pristine $VO_2$ film, indicating the recovery of the annealed $H_xVO_{2-\delta}$ sample. The Raman spectra in Fig.2 (b) showed the similar trend, which showed that $VO_{2-\delta}$ film was not able to recover to pure $VO_2$ by annealing at 200$^o$C, while by the same treatment, the H-doped $H_xVO_{2-\delta}$ sample could be transferred to pure $VO_2$ effectively. These results showed that hydrogen doping into $VO_{2-\delta}$ film could help to remove the oxygen vacancies in $VO_2$ crystal.

In order to carefully examine the dynamics of H-doped $H_xVO_{2-\delta}$ sample during the heat treatment, the in-situ XRD and Raman measurement was also conducted in Fig.2 (c) and (d). For the prepared $H_xVO_{2-\delta}$ sample, the XRD peak was at 36.3$^o$, which was quite consistent with the diffraction for fully hydrogenated $VO_2$ film [6,17,18]. While upon the annealing treatment in air at 200$^o$C, the XRD peak quickly shifted close to 39.8$^o$ within 5 minutes. With the longer annealing treatment up to five hours, the sample should recovered to pure $VO_2$ completely, since the XRD peak position was almost at 39.8$^o$ (See Fig. 3S). The in-situ Raman results in Fig. 2(d) were quite consistent with the XRD test, since after 5 minutes heat treatment, the Raman peaks for M-$VO_2$ phase were observed. The further heat treatment with longer time would make the complete removal of $V_O$ defects and the film transferred to pure $VO_2$ completely.

The synchrotron radiation based XANES tests for H-doped $H_xVO_{2-\delta}$ sample after the annealing treatment with different time were also conducted. From the curves in Fig.2 (e), the V L-edge curves shift continuously to higher energy as the $H_xVO_{2-\delta}$ film annealed in air (See Fig. 4S). This observation is quite understandable if considering the electron doping effect due to the hydrogen atoms insertion and the oxygen vacancies in the crystal lattice, which decreases the $V^{4+}$ valance state. By annealing treatment, the H atoms and oxygen vacancies were removed simultaneously, resulting the recovery of pure $VO_2$ film. Fig.2 (f) shows the related O K-edge curves for $H_xVO_{2-\delta}$ film before and after annealing in air. After the heated treatment, the relative intensity ratio of the $t_{2g}$ and $e_g$ peaks increased substantially, reflecting the variation of electron occupancy due to the removal of electron doping induced by the H atoms and oxygen vacancies.

To detect the doped H atoms and the oxygen vacancies in $VO_2$ crystal, the SIMS measurements for the depth distributions of H, V and O elements were conducted for the pristine $VO_2$ film, $VO_{2-\delta}$ film, $H_xVO_{2-\delta}$ film as well as the annealed samples, respectively. As shown in Fig.3 (a), the SIMS signal for hydrogen (H) in hydrogenated $H_xVO_{2-\delta}$ film was more than ten times larger than those for the pristine $VO_2$ and $VO_{2-\delta}$ film, indicating the insertion of H atoms into $VO_{2-\delta}$ film and proving that oxygen vacancies and H dopants can coexist stably. While for the heated $H_xVO_{2-\delta}$ film, the hydrogen (H) concentration decreased to the same level of pristine $VO_2$ and $VO_{2-\delta}$ film, which indicated the escaping of H atoms from $VO_2$ crystal. Fig.3 (b) shows the depth distributions of V element in different film sample. Almost the same SIMS signal were obtained, clarifying the fact that V atoms concentration was stable and no V related point defects were produced. Fig. 3 (c) showed the O element distributions, showing some difference for these tested samples. From the zooming SIMS depth profile of oxygen (O) element in Fig. 3 (d), it was observed that after heating treatment, the O atom concentration for $H_xVO_{2-\delta}$ film almost coincided with the pristine $VO_2$ film, indicating the removal of oxygen vacancies in the crystal film. These results suggested that under the heating conditions, H dopants were indeed helpful to remove the oxygen vacancies in $VO_{2-\delta}$ film. While if no H dopants was involved, the oxygen vacancies in $VO_{2-\delta}$ film was quite difficult to be removed even by the same heating treatment.

We have then carried out first-principles calculations to identify the underlying mechanism. Considering the position of H dopant in $VO_{2-x}$ cell, we built a $3\times3\times3$ $VO_2$ supercell model with one oxygen defect, and placed the hydrogen atom in all possible positions characterized by the distances from $V_O$ (**Table S1**). The most favorable position in thermodynamics among them is the point with a H-$V_O$ distance of 3.3 Å (Fig. 4d). A previous work has found that H dopants in metal oxide materials may promote the aggregation and migration of oxygen vacancies [13]. Therefore, we expect that in H-$VO_{2-x}$ the effusion of hydrogen dopants upon environmental heating could drive the removing of interior $V_O$ defects. Here we simulated the migration of $V_O$ in a $2\times1\times1$ $VO_{2-x}$ surpercell with or without H-doping, by using the Climbing-Image Nudge Elastic Band (CI-NEB) method to locate the transition states along $V_O$ migration pathway. The comparison of reaction profiles for H-$VO_{2-x}$ (Fig. 4a) and $VO_{2-x}$ (Fig. 4b) indicates that, with the help of H atom, the energy barrier of $V_O$ migration decreases by 0.46 eV. This thus supports the hypothesis of H dopant promoting $V_O$ diffusion. The hypothesis is also supported by simulated differential charge information in Fig. 4c, which shows appreciable accumulation of negative charges near $V_O$ in conjunction with the electron transfer from the H atom to V atoms near the oxygen vacancy. Such charge redistribution can weaken the V-O bond and thus enhance the mobility of O atom (equivalently, oxygen vacancy).

**Conclusion**

In conclusion, we have investigated the $V_O$ defects removal in $VO_2$ crystal film by annealing treatment at low temperature. It is observed that the metallic $VO_{2-\delta}$ crystal film is quite stable even by an annealing treatment at 200 ºC in the air. In the presence of hydrogen dopants, the $V_O$ can be effectively removed and pure $VO_2$ film with excellent MIT behavior is obtained. The theoretical simulations clarify that the H-doping induced electrons will accumulate around the $V_O$ sites and decrease the diffusion barrier energy, which promotes the easier removal of $V_O$ in $VO_2$ crystal. The current findings not only clarify the intrinsic interactions between the doped H atoms and the existed $V_O$ defects in crystal lattice, but also suggest a general approach to control the oxygen-related point defects in different oxide materials.

## Experimental Details

**Thin-film deposition and annealed conditions:** The 20-nm $VO_2$ (020) epitaxial films were grown on *c*-cut sapphire by an rf-plasma assisted oxide molecular beam epitaxy (rf-OMBE) equipment and more details for the film preparation are reported elsewhere [19]. To obtain oxygen-vacancies VO2 film samples named as $VO_{2-\delta}$, we annealed the pristine VO2 film in high vacuum ($p$~$1\times10^{-7}$ Torr) for 2 h at 350 ℃.

**Hydrogenation and heated treatment:** Before the hydrogenation treatment, a small amount of palladium nano-particles (~1 nm) were deposited onto the VO2 film surface by electron-beam evaporation at room temperature. Then the Pd-coated $VO_2$ samples were put into the horizon furnace and annealed in Ar/H2 mixed gas (15% H2) flow at 120 ℃ for 2h to form the heavily doped samples named as $HVO_{2-\delta}$. Then for removing the oxygen vacancies, both $VO_{2-\delta}$ and $HVO_{2-\delta}$ film samples were heat-treated in air at 200 °C with 5-300 min.

**Sample characterization:** The resistance-time and resistance-temperature electrical measurements were conducted by a customized four-probe system installed on a variable-temperature sample stage. The crystal structures of the pristine $VO_2$ film and $VO_{2-\delta}$ and $HVO_{2-\delta}$ film samples were characterized by X-ray diffraction (XRD, Model D/Max 2550 V, Rigaku, Japan) with Cu Kα radiation ($\lambda = 1.54178$Å) before and after the heating treament. Raman spectroscopy tests were recorded at room temperature with a LABRAM-HR Confocal Laser MicroRaman Spectrometer 750 K with a laser power of 0.5 mW. High sensitive secondary-ion mass spectrometry (SIMS) measurements (Quad PHI6600) were conducted to directly examine the hydrogen-vanadium-oxygen concentration in each sample. For each sample, we always conducted the above measurements at random multiple points, each test showed good consistency, confirming the uniformity of the as-deposited or heat-treated $VO_2$ samples.

**Synchrotron-based measurements**. Synchrotron X-ray diffraction spectra for the θ-2θ scanning was conducted at the BL14B1 beamline of the Shanghai Synchrotron

Radiation Facility (SSRF). The SSRF is a third-generate accelerator with a 3.5 GeV storage ring. The BL14B1 beamline shows the energy resolution (ΔE/E) of $1.5 \times 10^{-4}$ @10 keV and the beam size of $0.3 \times 0.35$ mm with the photo flux of up to $2 \times 10^{12}$ phs/s@10 keV. Considering the photo flux distribution and the resolution, the 0.12398 nm X-ray was chosen during the experiment. The X-ray absorption near-edge spectroscopy (XANES) was conducted at the XMCD beamline (BL12B) in National synchrotron radiation laboratory (NSRL), Hefei. The total electron yield (TEY) mode was applied to collect the sample drain current under a vacuum better than $3.75 \times 10^{-10}$ Torr. The energy range is 100–1000 eV with an energy resolution of 0.2 eV.

**First-principles calculations:**

To elucidate the migration pathways of oxygen vacancies in H-doped $VO_{2-x}$, we performed spin-polarized density functional theory (DFT) calculations using the Perdew−Burke−Ernzerhof (PBE)[20] functional and the plane-wave projector augmented wave (PAW)[21] method as implemented in the Vienna ab initio simulation package (VASP)[22]. The energy cutoff for the plane-wave basis set is 450 eV. The convergence criteria for structure relaxation are $10^{-5}$ eV on the energy and 0.01 eV/Å on the residual force of each atom. The Brillouin zone was sampled with a Monkhorst-Pack scheme[23] with $3 \times 3 \times 6$ k-point grids.

**Acknowledgements**

This work was partially supported by the National Key Research and Development Program of China (2016YFA0401004), the National Natural Science Foundation of China (11574279, 11704362), the funding supported by the Youth Innovation Promotion Association CAS, the Major/Innovative Program of Development Foundation of Hefei Center for Physical Science and Technology and the China Postdoctoral Science Foundation (2017M622002). This work was partially carried out at the USTC Center for Micro and Nanoscale Research and Fabrication. The authors also acknowledged the supports from the Anhui Laboratory of Advanced Photon Science and Technology. The approved beamtime on the XMCD beamline (BL12B) in

National Synchrotron Radiation Laboratory (NSRL) of Hefei was also appreciated.## References

[1] F. J. Morin, Physical Review Letters **3**, 34 (1959).
[2] Z. Yang, C. Ko, and S. Ramanathan, Annual Review of Materials Research **41**, 337 (2011).
[3] M. Liu *et al.*, Nature **487**, 345 (2012).
[4] J. Del Valle *et al.*, Nature **569**, 388 (2019).
[5] W. Yi, K. K. Tsang, S. K. Lam, X. Bai, J. A. Crowell, and E. A. Flores, Nat Commun **9**, 4661 (2018).
[6] S. Chen, Z. Wang, H. Ren, Y. Chen, W. Yan, C. Wang, B. Li, J. Jiang, and C. Zou, Sci Adv **5**, eaav6815 (2019).
[7] J. Jeong, N. Aetukuri, T. Graf, T. D. Schladt, M. G. Samant, and S. S. P. Parkin, Science **339**, 1402 (2013).
[8] Y. Sharma *et al.*, ACS Nano  (2018).
[9] K. Appavoo, D. Y. Lei, Y. Sonnefraud, B. Wang, S. T. Pantelides, S. A. Maier, and R. F. Haglund, Jr., Nano Lett **12**, 780 (2012).
[10] Z. Zhang *et al.*, Physical Review Applied **7** (2017).
[11] H.-T. Zhang *et al.*, Adv. Funct. Mater. **26**, 6612 (2016).
[12] T. L. Meyer, R. Jacobs, D. Lee, L. Jiang, J. W. Freeland, C. Sohn, T. Egami, D. Morgan, and H. N. Lee, Nat Commun **9**, 92 (2018).
[13] L. Luo *et al.*, Nat Mater  (2018).
[14] T. Kamiya, K. Nomura, and H. Hosono, Journal of Display Technology **5**, 468 (2009).
[15] J. Kim, S. Bang, S. Lee, S. Shin, J. Park, H. Seo, and H. Jeon, Journal of Materials Research **27**, 2318 (2012).
[16] C. Chen *et al.*, Adv Sci (Weinh) **6**, 1801189 (2019).
[17] S. Chen *et al.*, Physical Review B **96**, 125130 (2017).
[18] H. Yoon, M. Choi, T. W. Lim, H. Kwon, K. Ihm, J. K. Kim, S. Y. Choi, and J. Son, Nat Mater **15**, 1113 (2016).
[19] L. L. Fan, S. Chen, Y. F. Wu, F. H. Chen, W. S. Chu, X. Chen, C. W. Zou, and Z. Y. Wu, Applied Physics Letters **103**, 131914 (2013).
[20] J. P. Perdew, K. Burke, and M. Ernzerhof, Phys. Rev. Lett. **77 (18)** (1996).
[21] P. E. Blochl, Phys Rev B Condens Matter **50**, 17953 (1994).
[22] G. Kresse and J. Furthmüller, Comput. Mater. Sci. **6 (1), 15-50** (1996).
[23] H. J. Monkhorst and J. D. Pack, Physical Review B **13**, 5188 (1976).
10

**Figures and Captions**

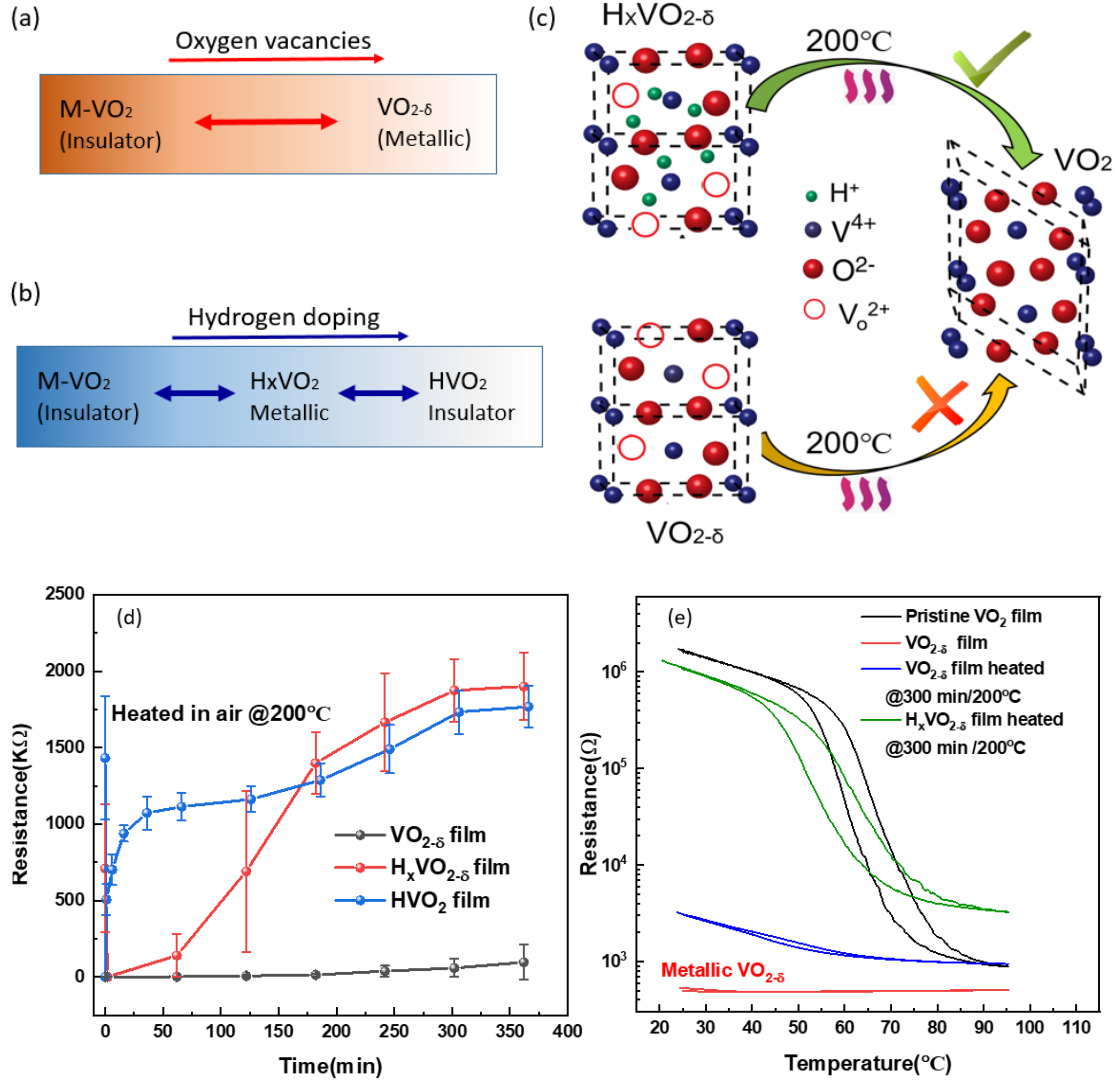

**Fig.1** (a) The transformation between the pristine $VO_2$ insulator and the metallic $VO_{2-\delta}$. (b) The scheme for H-doped $VO_2$, which shows three phase-transitions due to the different H-doping concentration. (c) The schematic removal of oxygen vacancies in $VO_{2-\delta}$ samples. Annealing the $VO_{2-\delta}$ sample in air at 200°C cannot transfer it to pure $VO_2$, while hydrogen doping sample can be recovered to pure $VO_2$ effectively. (d)The electrical resistance as the function of annealing time for $VO_{2-\delta}$, $H_xVO_{2-\delta}$ and $HVO_2$ samples by heating in air at 200°C; (e) The R-T curves for the pristine $VO_2$ film, the metallic $VO_{2-\delta}$ before and after heating treatment and the $H_xVO_{2-\delta}$ sample after heating treatment, respectively.

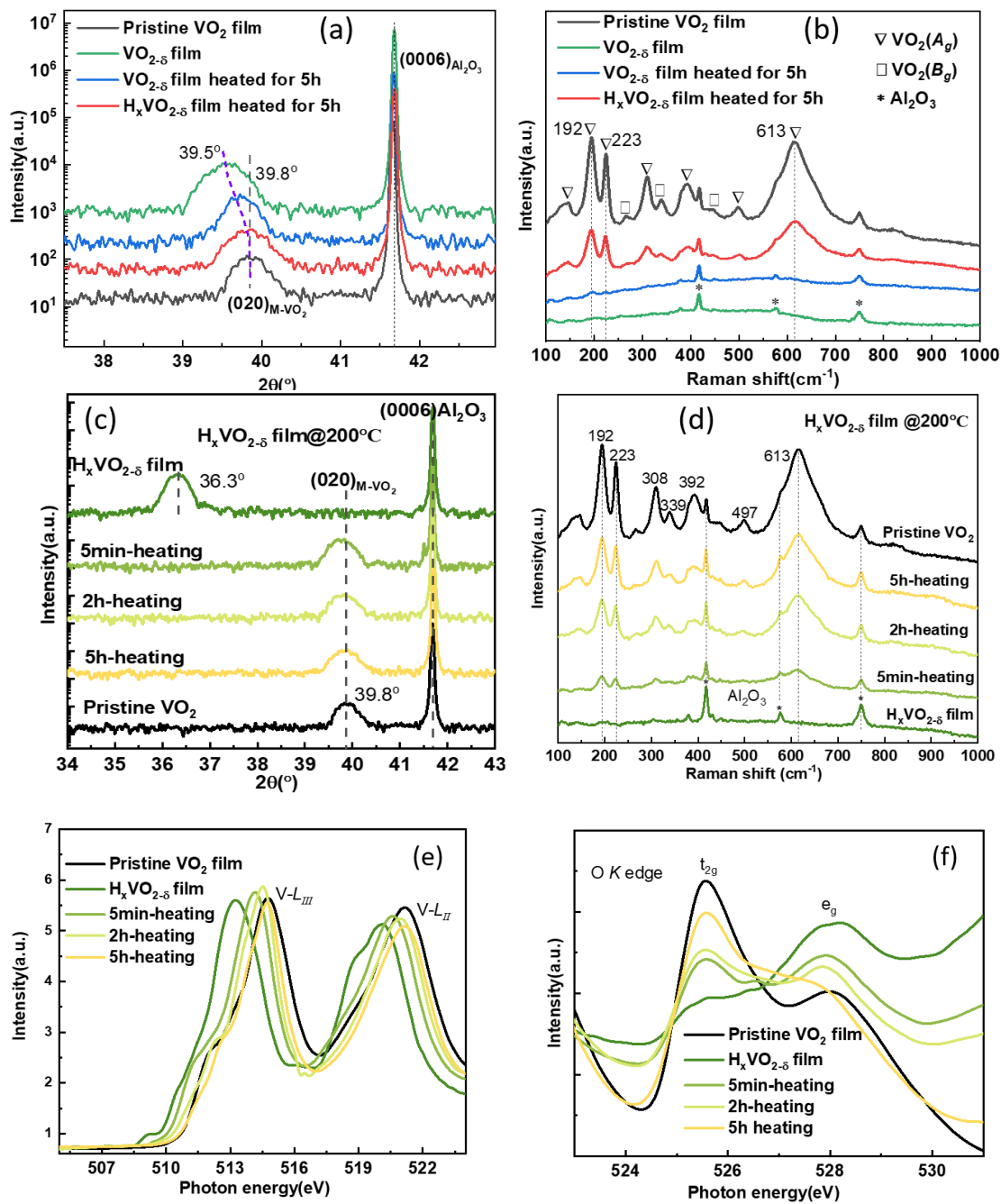

**Fig.2** (a~b) The XRD and Raman spectra for the pristine $VO_2$ film, $VO_{2-\delta}$ film and the 5-hour annealed $VO_{2-\delta}$ and $H_xVO_{2-\delta}$ film samples; (c~f) The XRD, Raman and XANES measurements for pristine $VO_2$ and $H_xVO_{2-\delta}$ film with the heating time of 5min, 2h and 5h, respectively.

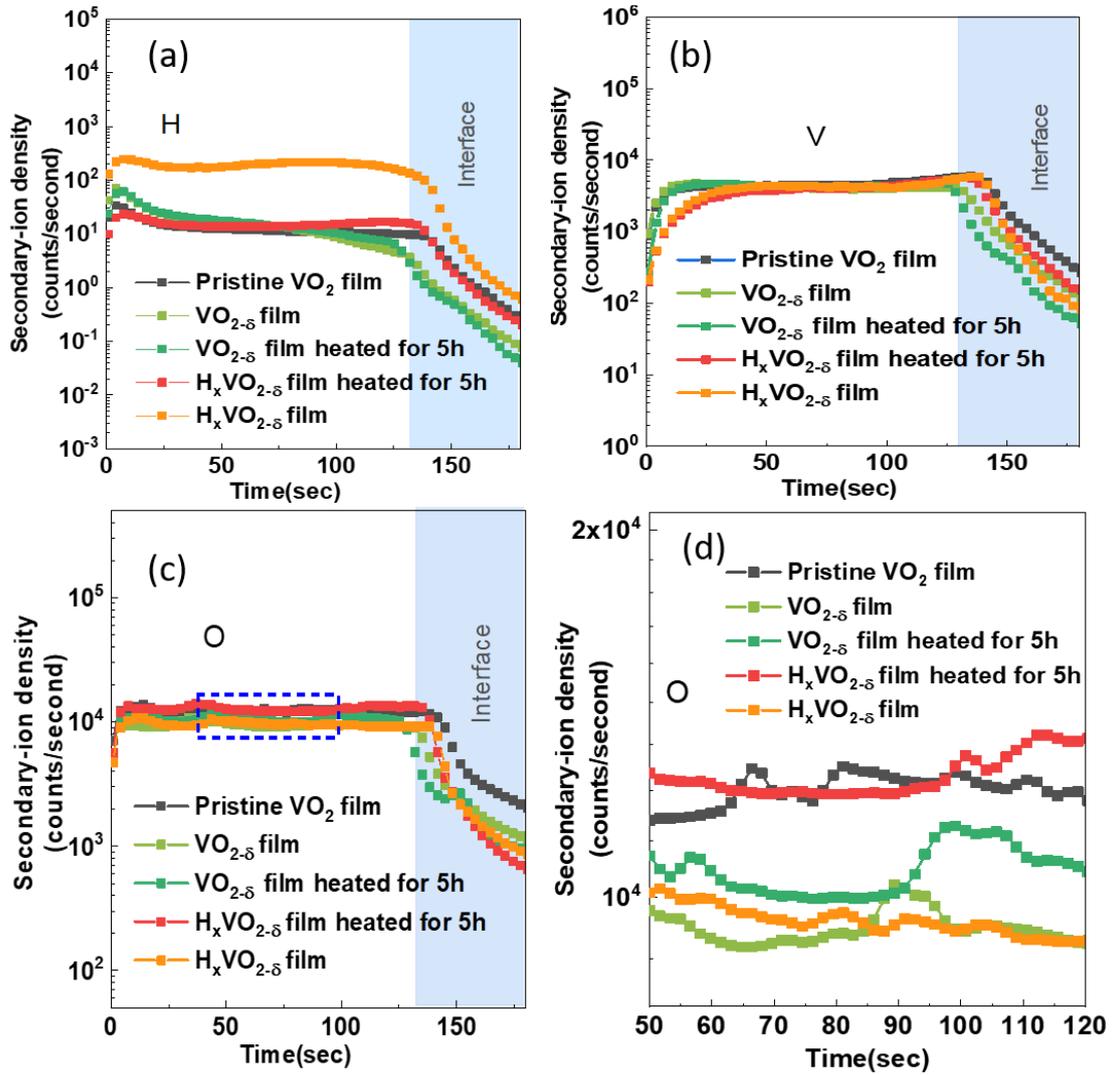

**Fig. 3** (a~c) SIMS depth profiles of H, V and O elements for the pristine $VO_2$ film, $VO_{2-\delta}$ film, $H_xVO_{2-\delta}$ film and the 5-hour annealed $VO_{2-\delta}$ and $H_xVO_{2-\delta}$ film samples. (d) The enlarged SIMS depth profiles of O element in Fig. 3c (the blue rectangle area), which clearly shows the variation of O element concentration in different samples.

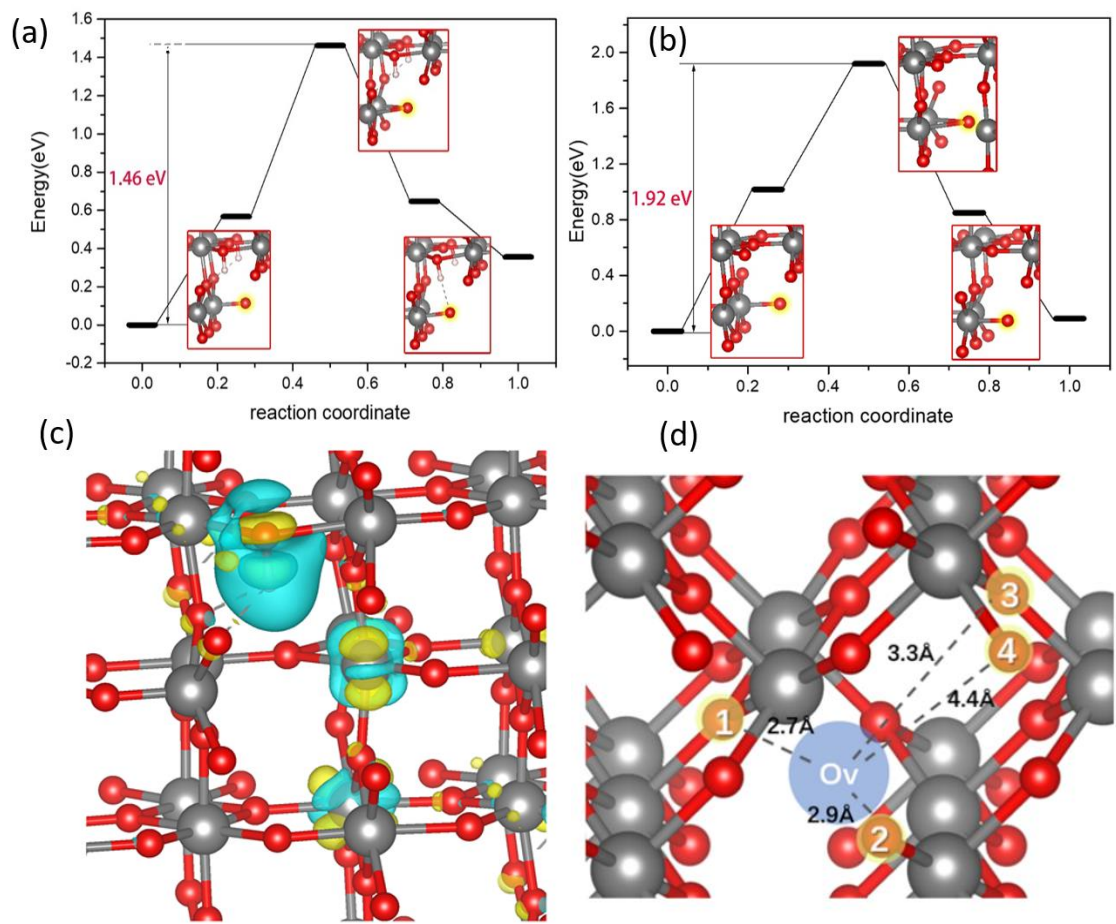

**Fig. 4.** The energy landscape and charge density difference of VO$_{2-X}$ supercell. (a) The energy landscape with H atom doping. (b) The energy landscape without H atom doping. (c) The charge density difference of VO$_{2-X}$ with H atom doping. (d) The most thermodynamically favorable position of H dopant in VO$_{2-X}$ cell.